# Cross-linking of polyolefins : a study by thermoporosimetry with benzene derivatives swelling solvents

*Nicolas Billamboz[1], Jean-Marie Nedelec[2], Manuel Grivet[1] and Mohamed Baba[3]\*.*

[1] Laboratoire de Microanalyses Nucléaires Alain Chambaudet, UMR CEA, Université de Franche-Comté, 16 route de Gray 25030 Besançon Cedex France.

[2]Laboratoire des Matériaux Inorganiques, UMR CNRS 6002,

Ecole Nationale Supérieure de Chimie de Clermont-Ferrand et Université Blaise Pascal

24, avenue des Landais 63174 Aubière Cedex France.

[3] Laboratoire de Photochimie Moléculaire et Macromoléculaire, UMR CNRS 6505,

Ecole Nationale Supérieure de Chimie de Clermont-Ferrand et Université Blaise Pascal

24, avenue des Landais 63174 Aubière Cedex France

\*Corresponding Author
e-mail: **mohamed.baba@univ-bpclermont.fr**
Phone: 00 33 (0)4 73 40 71 61
Fax: 00 33 (0)4 73 40 77 00

ABSTRACT

$o$, $m$, $p$-xylene, $p$-dichlorobenzene, 1,2,4 trichlorobenzene and naphthalene were calibrated as condensates used in thermoporosimetry technique. Exponential relationships were found connecting the pore radii ($R_p$(nm)) and the freezing point depression of the swelling solvent ($\Delta T$ (C°)) on one side and the apparent energy of crystallization ($W_a$ (J.cm$^{-3}$)) and $\Delta T$ on the other side:

$$R_p = t.\exp(-\frac{1}{c.\Delta T})$$
$$W_a = W_0.\exp(\frac{\Delta T}{f})$$

Pore or mesh size distribution can be derived from DSC thermal recording using the following equation:

$$\frac{dV_p}{dR_p} = k.\frac{c.Y(T).\Delta T^2}{W_a.R_p}$$

All the numerical parameter values were determined.

Polyethylene and polypropylene samples, cross-linked with high-energy electrons or γ-rays, were submitted to thermoporosimetry study. Relative mesh size distributions (MSD), depending on polymer/solvent pair, were calculated for these polyolefins, using $o$, $m$ and $p$-xylene as solvent and were found in the same sequences that their degrees of swelling and the irradiation doses they received.

# Introduction

Most of polymers are used under cross-linked form. Cross-linking is known as the common mean to enhance both the mechanical and the rheological properties of polymeric materials. The study of cross linking is difficult since once cross linked, the polymer is no longer soluble. Several techniques are devoted to quantify the degree of cross-linking of polymer: gel fraction [1], measurement of relaxation time [2,3] or mechanical tests, but all of them only give average values.

In this paper, we aim to expose the results of the thermal analysis approach allowing calculation of mesh size distribution of the cross-linked network of polyolefins such as polyethylene and polypropylene. The paper proposes an application of the thermoporosimetry technique to swollen polymeric gels using varying solvents such as xylene (*o, m and p*), *p*-dichlorobenzene and naphthalene. Thermoporosimetry is a textural analysis technique based on the measurement of the shift ($\Delta T$) of the temperature of the thermal transition undergone by a liquid when it is confined inside a divided medium. Porous media like silica and alumina were the first materials investigated with this technique [4]. Their pore size distributions (PSD) [6,7] and their porous volume ($V_p$) [8] were calculated as precisely as it can be done by gas sorption or mercury porosimetry. Solvent, inside a polymeric swollen gel, behaves like a liquid trapped inside a porous material, notably its thermal transitions (liquid to solid and also solid to solid) are shifted in the same way as affected in porous media. Both hydro [9] and elastomeric [10,11] swollen gels were successfully studied with this technique. For these organic materials, the polymeric network is assumed to be built by a collection of adjacent cells or meshes whose size distribution (MSD) characterizes the cross-linking level.

Numerous theoretical development have been performed by Brun and co-workers [4] extending the Gibbs and Thomson's works. According to those authors, there is a direct relationship between the size of the cavity or of the zone where the liquid is "compartimentized" [5] ($R_p$) and the inverse of the shift of its thermal transition temperature ($\Delta T$):

$$R_p = f(1/\Delta T) \qquad (1)$$

$$\Delta T = T_p - T_0 \qquad (2)$$

where $T_0$ is the temperature of the thermal transition of the bulk liquid and $T_p$ is that of the confined one.

In the water and benzene case, Brun et al. [4] propose a linear relationship between $R_p$ and $\Delta T^{-1}$ whereas *n*-hexane, *n*-tridecane, *n*-tetradecane and *n*-tricosane [12] show a non-linear behavior.

When a confined solvent, soaking porous rigid material or swelling polymeric cross-linked gel, is submitted to DSC analysis with a decreasing temperature program, its recorded thermal curve, called in this work "thermogram", exhibits two peaks: the first one is attributed to the crystallization of the bulk liquid and the second to the confined one. Each point of the peak of confined solvent can be related to a specific $R_p$ value. Consequently, the thermogram just reflects, in a first approximation, the pore parameters. However, to derive the right size distribution, it is necessary to take into account the decrease of the thermal transition energy when $\Delta T$ increases. In fact, the elementary volume ($dV_p$) of the confined liquid affected by the thermal transition can be estimated as follows [4] :

$$dV_p = k \cdot \frac{Y(T) \cdot d(\Delta T)}{W_a} \qquad (3)$$

where $dV_p$ is the elementary volume of the confined solvent which undergoes crystallization, $Y(T)$ the thermogram ordinate, $\Delta T$ the shift of the transition temperature, $W_a$ is the apparent energy taking into account the thermal dependence of heat of crystallization and the amount of solvent remaining adsorbed and which does not take part in the transition and $k$ is a proportionality constant.

Pore or mesh size distributions can be calculated as following [4]:

$$\frac{dV_p}{dR_p} = f(R_p) \qquad (4)$$

To our knowledge, only few solvents including water, benzene, cyclohexane, carbon tetrachloride and *n*-heptane have been investigated as condensate in thermoporosimetry measurements. Since these solvents are not suitable to swell polyolefins (in particular polyethylene and polypropylene), we decided to calibrate some more appropriate solvents including o,m,p-xylene, p-dichlorobenzene, 1,2,4 trichlorobenzene and naphthalene. The numerical relationships, resulting from this calibration, allow a practical calculation of mesh size distributions of those polymers from an easy DSC experiment.

Two types of numerical relationships must be established: the first one is represented by Equation (1) and the second one giving $W_a = f(T)$. We choose a direct calibration approach using porous silica samples with well known and narrow pore size distributions. The pore size distributions and the porous volumes of those silica samples have been measured by gas sorption method.

## Materials and Methods

All organic compounds (o,m,p-xylene, p-dichlorobenzene, 1,2,4 trichlorobenzene and naphthalene) are supplied by Aldrich company and they are of HPLC quality and were used without further purification.

The porous gel-derived silica monoliths (2.5 mm × 5.6 mm diameter cylinders) were prepared by the acid catalysed hydrolysis and condensation of an alkoxysilane precursor, following the procedures described in [13]. They are the same samples already used in precedent works [10,11].

Careful control of the ageing procedure performed at 900°C allowed the production of matrices with tailored textural properties. In this study five different samples corresponding to different ageing procedures were used. The textural properties (Specific Surface Area (SSA), total pore volume ($V_p$) and Pore Size Distribution (PSD)) were determined by $N_2$ sorption on a Quantachrome AS6 Autosorb apparatus. Five samples with different pore sizes were used. Their textural data are displayed in Table 1. In this table, the modal pore radius corresponding to the maximum of the PSD (MaxPSD) is also given. This value was used for the DSC calibration.

| Sample | SSA ($m^2.g^{-1}$) | $\sigma_{SSA}$ ($m^2.g^{-1}$) | $V_p$ ($cm^3.g^{-1}$) | $\sigma_{Vp}$ ($cm^3.g^{-1}$) | Mean $R_p$ (nm) | $\sigma_{Rp}$ (nm) | MaxPSD $R_p$ (nm) |
|---|---|---|---|---|---|---|---|
| 1.25 nm | 537.0 | 12.8 | 0.333 | 0.001 | 1.24 | 0.05 | 1.75 |
| 2.5 nm | 532.0 | 9.5 | 0.696 | 0.028 | 2.62 | 0.30 | 2.40 |
| 3.75 nm | 472.7 | 1.5 | 0.922 | 0.082 | 3.90 | 0.67 | 3.42 |
| 10.0 nm | 166.2 | 3.2 | 0.991 | 0.071 | 11.92 | 1.24 | 8.70 |
| 13.5 nm | 183.1 | 1.0 | 1.327 | 0.072 | 14.49 | 1.41 | 14.25 |

Table 1: Textural data of the porous gel-derived silica samples and their corresponding deviations.

To validate the calibration, a control alumina sample supplied by Coulter Company was used and its textural data (S=212.8 $m^2.g^{-1}$, $V_p$ = 0.510 $cm^3.g^{-1}$) were determined on a Coulter BET SA3100 apparatus.

The polyethylene used in this study is an industrial polymer of linear medium density polyethylene (density = 0.935 $g.cm^{-3}$), with narrow molecular distribution. The additives represent about 1% of the total weight. The cross-linking of the samples has been performed by irradiation with high-energy electrons. Both sides of the samples (6 mm thick) have been irradiated with 2.2 MeV electrons with a Van de Graaff apparatus up to 1000 kGy. The homogeneity of irradiation has been verified by a Monte Carlo simulation code: MCNP. The PP/EPR is also an industrial polymer based on the mixture of a polypropylene homopolymer, an ethylene-propylene copolymer and an high-density polyethylene (HDPE). EPR surrounds high-density polyethylene (HDPE), and the dispersed particles of HDPE and EPR (the diameter varies from 1 to 4 μm) are submerged in PP. The HDPE-EPR nodules represents about 15% (by mass) of PP/EPR, and each nodule is composed of about two-thirds HDPE and one-third EPR. The samples of PP/EPR have been irradiated by γ-rays from $^{60}$Co (1.17 MeV and 1.33 MeV). Table 2 gives the dose and swelling ratio, values related to all those polyolefins. The swelling ratio is defined as the ratio between the volume of the swollen gel and the volume of the dry gel.

| Samples | Doses (kGy) | Swelling ratio $p$-xylene 138°C | Swelling ratio $m$-xylene 138°C | Swelling ratio $o$-xylene 138°C |
|---|---|---|---|---|
| PE1000kGy | 1000 (electrons) | 3.86 | 3.72 | 4.46 |
| PP940kGy | 940 (γ) | 9.01 | 8.86 | 8.78 |
| PP830kGy | 830 (γ) | 11.45 | 9.54 | 9.70 |
| PP620kGy | 620 (γ) | 13.78 | 12.21 | 12.28 |
| PP405kGy | 405 (γ) | 25.78 | 20.93 | 23.84 |

Table 2: Doses of irradiation and swelling ratios of the studied polyolefins.

A Mettler Toledo DSC 30 and a TA Instrument 2920 MDSC apparatus are used to perform thermal analysis. They are calibrated with indium and zinc reference samples and are equipped with a liquid nitrogen cooling system allowing to work in the temperature range between −150°C and 600°C. Dedicated software is used to calculate temperatures and transition heats from the thermal DSC curves.

**Results and discussion**

**Calibration**

As stated in the introduction part, it is necessary to establish numerical relationships between pore radii ($R_p$) and the shift of crystallization temperature ($\Delta T$) on one hand and between the apparent energy ($W_a$) and $\Delta T$ on the other hand. Due to the lack of available data related to the organic compounds, we choose to use a direct calibration approach to reach these relationships. Silica gel samples have very well defined textural properties namely the average pore radius ($R_p$) and the porous volume ($V_p$). The knowledge of the $R_p$ values and the experimental determination of corresponding $\Delta T$ are enough to establish the relation (1) while the porous volume and the measurement of the heat of crystallization of the confined solvent at the concerned temperature allow us to reach $W_a$.

A monolithic silica gel, weighing around 40 mg, is set in DSC 140 µL pan and a few solvent drops are added. A visual control, following the spread of the liquid through the sample, is enough to make sure that, after 20 minutes, the silica is completely imbibed. Some extra drops are then added in order to maintain the bulk solvent in a slight excess. Naphthalene and p-dichlorobenzene are under solid state at room temperature. So, after setting the silica pastille in the bottom of the DSC pan, the later is filled with naphthalene or p-dichlorobenzene crystals prior staying for 20 minutes in a thermally regulated oven maintained at 85°C. The crystals are added until the bulk liquid is in excess and then the DSC pan is sealed and introduced in the apparatus for thermal analysis. A 0.7°C.min$^{-1}$ cooling rate was chosen. This cooling rate is slow enough to allow the continuous thermal equilibrium inside the DSC cell.

As shown in a subsequent heating (Figure 1 (**i**)) of the sample from −80°C up to −10°C, two endothermic peaks appear: the first one is attributable to the confined *o*-xylene and the second one reveals the fusion of the bulk solvent. To avoid the initial super-cooling, a heating (Figure 1 : (**b**)) is then run from −80°C and stopped at −24°C just after the confined solvent was melted. At this temperature, the confined solvent is still liquid but in close contact with the crystallites corresponding to the bulk solvent. These crystallites serve as nucleation sites and the crystallisation of the confined solvent upon further cooling (Figure 1 : (**c**)) down to -80 °C, is facilitated and supercooling effect is then avoided.

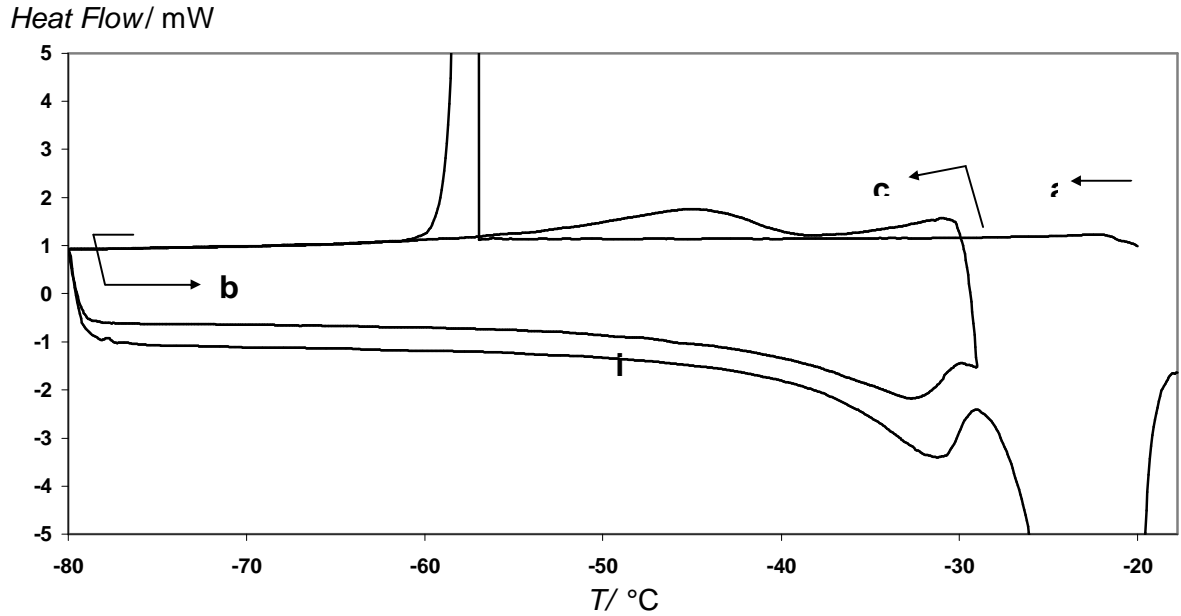

Figure 1: DSC thermograms of PE sample electron-irradiated at 1000 kGy and swollen with o-xylene. a: the first cooling, showing a dramatic super-cooling phenomenon, b: the first heating is stopped just before the bulk solvent was melted and c: the second cooling showing the peak of the trapped o-xylene inside the polymeric gel. The "i" thermogram is also presented in order to show the two endothermic peaks.

Two kinds of data can be derived from the (c) DCS thermal curve: the peak position temperature ($T_p$) leading to the $\Delta T$ calculation and the area under the peak ($\Delta H$ (J)) permitting to calculate the apparent energy ($W_a$ ( J.cm$^{-3}$)) as follows:

$$W_a = \frac{\Delta H}{m.V_p} \qquad (5)$$

where $m$ (g) is the weight of the dry silica gel sample and $V_p$(cm$^3$.g$^{-1}$) its porous volume.

For the smaller $R_p$ values (1.25, 2.5 and 3.75 nm samples), the confined peak is beyond the breaking super-cooling peak and a direct cooling is enough.

Figure 2 gives the six thermograms series in relation to the six organic compounds under study. For *o* and *m*-xylene no signal is detected for the 1.25 nm sample; it may crystallize at a lower temperature than −150°C or it could be that its $W_a$ value is too weak to be recorded.

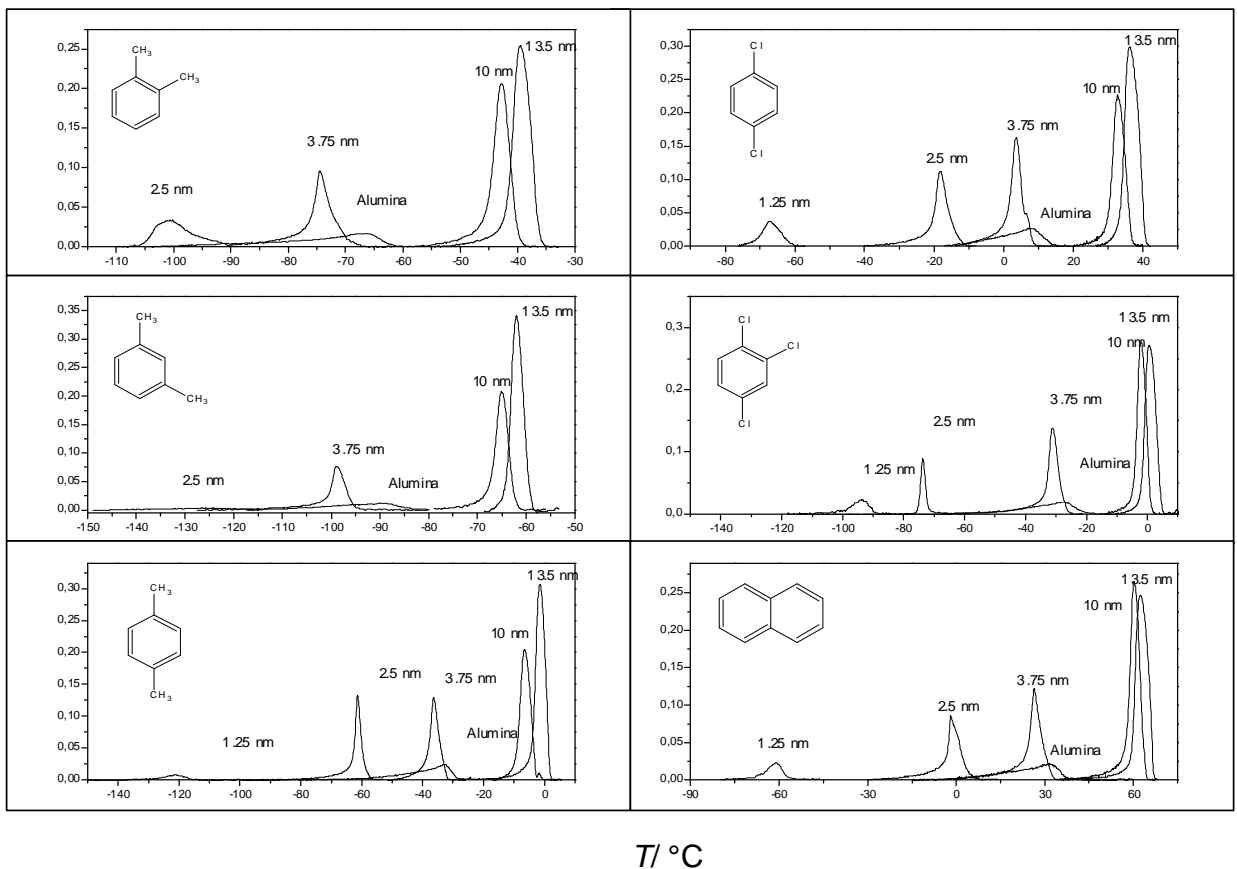

**Figure 2**: Thermograms of the six organic compounds (*o, m, p*-xylene, *p*-dichlorobenzene, *1,2,4*-trichlorobenzene and naphthalene) soaking the silica gel samples referenced 13.5, 10, 3.75, 2.5 and 1.25 nm (see table 1). The thermograms of the six solvents confined inside the alumina test sample are also presented. Only the peaks of confined solvent are shown. The decreasing temperature rate is 0.7 °C.min$^{-1}$.

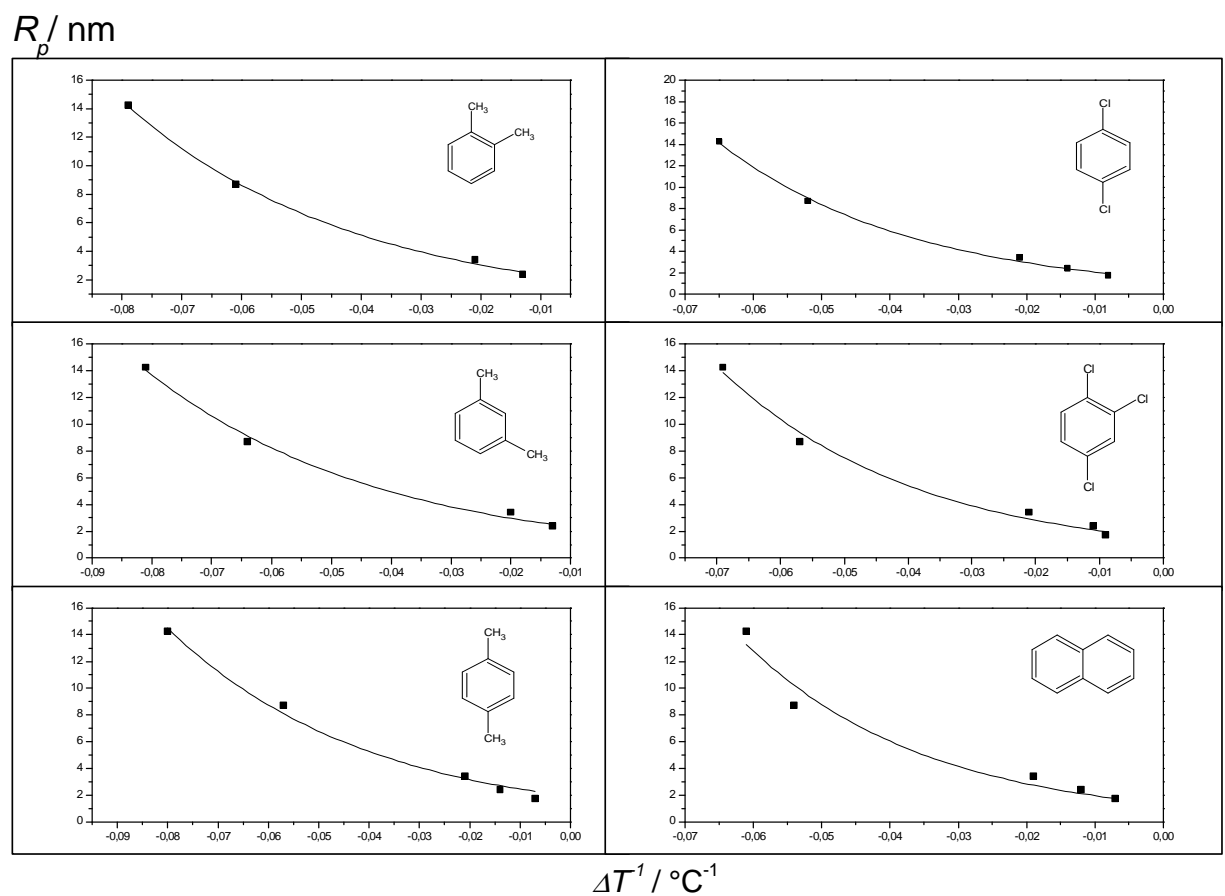

**Figure 3**: Pore radii ($R_p$) of the silica gel calibration samples *vs* ($\Delta T^{-1}$) for each of the six solvents (o, m, p-xylene, p-dichlorobenzene, 1,2,4-trichlorobenzene and naphthalene).

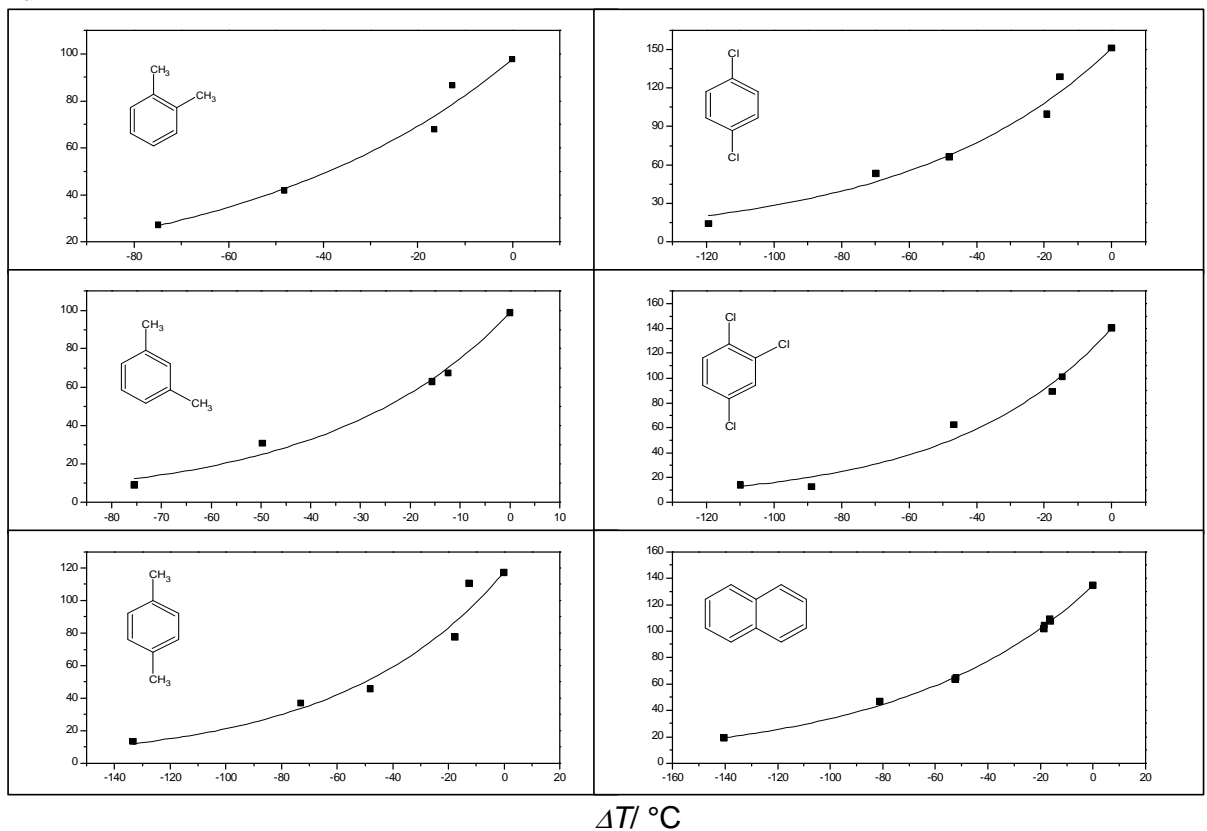

**Figure 4**: Plots of apparent energy ($W_a$) of crystallization *vs* freezing temperature depression *(ΔT)* of the confined solvents. This change in the thermal transition energy takes into account two contributions: the first is the decreasing of latent enthalpy when the temperature decreases and the second is the fact of an amount of the confined solvent remains adsorbed on the wall of the divided medium and does not take part to the thermal transition.

Figure 3 and 4 show the plots of $R_p$ versus $ΔT^{-1}$ and $W_a$ versus $ΔT$ respectively. The best fits of these experimental results give exponential dependences between both $R_p(nm)$ and $ΔT^{-1}$ on one hand and $W_a$ (J.cm$^{-3}$) and $ΔT$ on the other hand. The following relationships were established:

$$R_p = t \cdot \exp(-\frac{1}{c \cdot \Delta T}) \qquad (6)$$

$$W_a = W_0 \cdot \exp(\frac{\Delta T}{f}) \qquad (7)$$

$c$ and $f$ are parameters depending on the solvent. $W_0$ and $t$ are also constant. $W_0$ is the energy of crystallization of the bulk solvent at its normal solidification temperature and $t$ being the thickness of the layer of adsorbed solvent.

|  | $R_p = t \cdot \exp(-\frac{1}{c \cdot \Delta T})$ | | | $W_a = W_0 \cdot \exp(\frac{\Delta T}{f})$ | | |
|---|---|---|---|---|---|---|
|  | t (nm) | c (°C$^{-1}$) | R$^2$ | W$_0$ (J/cm$^3$) | f (°C) | R$^2$ |
| o-xylene | 1.814 | 0.03842 | 0.998 | 97.7 | 57.9 | 0.971 |
| m-xylene | 1.789 | 0.03930 | 0.995 | 98.7 | 36.6 | 0.989 |
| p-xylene | 1.909 | 0.03948 | 0.992 | 117.2 | 58.1 | 0.928 |
| p-dichlorobenzene | 1.462 | 0.02860 | 0.998 | 151.3 | 59.7 | 0.975 |
| 1,2,4-trichlorobenzene | 1.474 | 0.03076 | 0.991 | 140.3 | 46.1 | 0.974 |
| Naphthalene | 1.345 | 0.02660 | 0.967 | 134.5 | 73.7 | 0.999 |

**Table 3**: Numerical parameter values of the relationships between the apparent energy of crystallization ($W_a$) and the freezing point depression ($\Delta T$) on the one hand and the pore radii ($R_p$) and ($\Delta T^{-1}$) one the other hand.

Table 3 summarizes the numerical values of the calculated parameters in relation to the six organic compounds.

Equation (6) gives the dependence between the size ($R_p$) of the cavity or the zone in which the solvent is confined and the shift ($\Delta T$) of its temperature of crystallization. For all the studied solvents, a general exponential relationship was found. A physical meaning can be attributed to the $t$ parameter if one remarks that when $\Delta T$ tends towards the infinite, $R_p$ takes the $t$ value. $t$ can then be assimilated to the thickness of the layer of adsorbed solvent on the internal wall of the pores. For water, benzene and *n*-heptane this thickness was found [4,10] ranging from 2 to 6 nm. For the six solvents under study, $t$ is ranging between 1.3 and 1.9 nm. Besides, a linear relationship like that established for the three solvent mentioned above can be derived from the Equation (6) for the high $\Delta T$ values. In fact, the first order Taylor polynomial gives:

$$R_p \approx -\frac{t}{c.\Delta T} + t \qquad (8)$$

The calculated $t/c$ values, derived from Table 2, were found around 50 nm.K that is in an acceptable agreement with 131.6, 64.7[4] and 58.7 nm.K [10] values published for benzene, water and *n*-heptane respectively.

Equation (7) gives the evolution of the heat of crystallization with temperature. $W_a$, apparent energy, takes also into account the amount of the solvent remaining adsorbed and being not affected by the change of state. $W_0$ is the value of $W_a$ related to the bulk solvent, when $\Delta T$ equals zero.

Pore or mesh size distribution then can be calculated by use of Equation (4) which takes the following analytic expression:

$$\frac{dV_p}{dR_p} = k . \frac{c.Y(T).\Delta T^2}{W_a.R_p} \qquad (9)$$

To check the consistency of these calibration results, an alumina porous sample, whose PSD was previously calculated by BJH gas adsorption method [14], is submitted to thermoporosimetry analysis. Its crystallization DSC curve is recorded in the same conditions as used for the silica calibrating samples.

Thermograms of this alumina control sample are superimposed on those of silica gel in figure 2. It is remarkable that, whatever the solvent, the thermogram of the alumina control appears just before that of the 3.75 nm silica sample.

After the base line was subtracted from the DSC curve, the latter is transformed, using Equation (9), into PSDs as can be seen in Figure 5. A good agreement is observed between the thermoporosimetry PSDs and that calculated from BJH method.

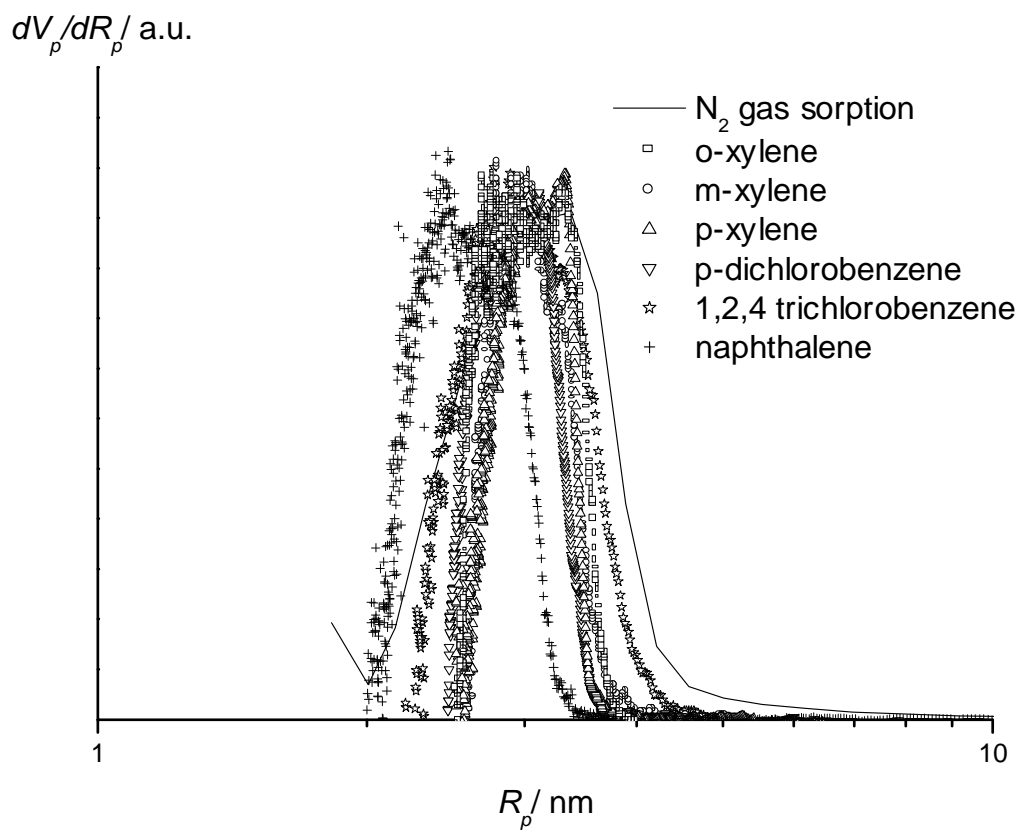

**Figure 5**: Pore size distributions of the alumina test sample calculated from the $N_2$ sorption technique (line) compared with that derived from thermoporosimetry calculation (symbols) using each of the six solvents (o, *m,,p*-xylene, *p*-dichlorobenzene, *1,2,4*-trichlorobenzene and naphthalene) as condensate.

**Evaluation of the cross-linking of polyolefinic polymers**

The solvent inside swollen polymer is in strong interaction with the macromolecular network. It is "compartimentized" and its degrees of freedom is considerably limited. Thus a formal analogy can be done between this solvent and a liquid confined in a porous material even if there are no pores in the gel. Polyethylene (PE) and Polypropylene/Ethylene-Propylene Rubber (PP/EPR) were soluble in all the six organic compounds at high temperature (138°C). The *1,2,4*- trichlorobenzene (TCB), for example, is a current solvent of these polyolefins for the SEC (Size Exclusion Chromatography) analysis. When PE or PP/EPR is cross-linked they become insoluble in these solvents but these latter can still swell them especially at high temperature.

About 10 mg of swollen polymer sample is introduced in a DSC pan and submitted to the same procedure already described in the calibration section. The swelling is performed in a reflux assembly maintained at 138°C thanks to a silicon thermo-regulated bath. Given the size of the sample (10 mg), the maximum swelling is reached after refluxing for one or two hours. After the swelling at 138°C, the reflux is stopped and the sample submitted to DSC analysis. As the swollen sample undergoes a cooling when it is transferred from the reflux assembly to the DSC apparatus, it is necessary to observe how its swelling equilibrium is modified. Figure 6 gives three thermograms of the same PE sample swollen with *o*-xylene: the first one is performed immediately after the reflux was stopped, the second one after two hours resting at room temperature (20°C) and the third after one week at room temperature. The swelling state of the sample seems to be stabilized after a few hours at 20°C.

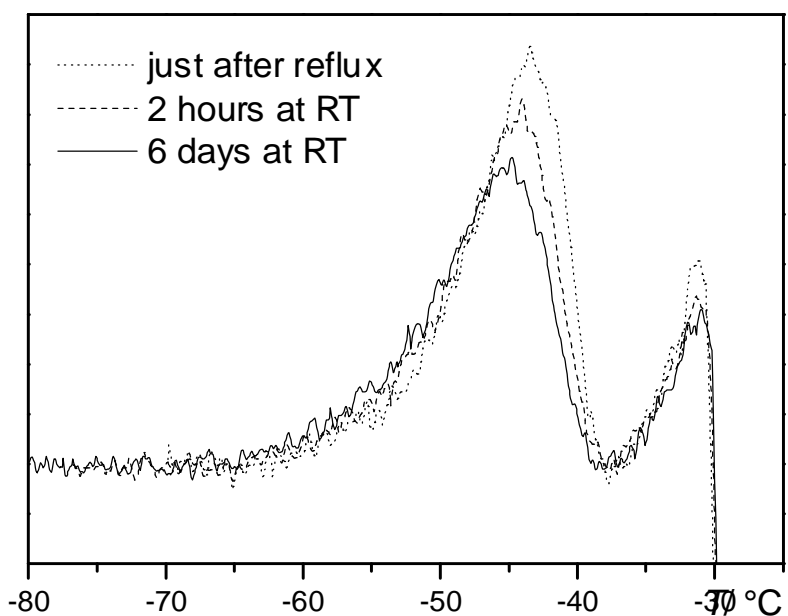

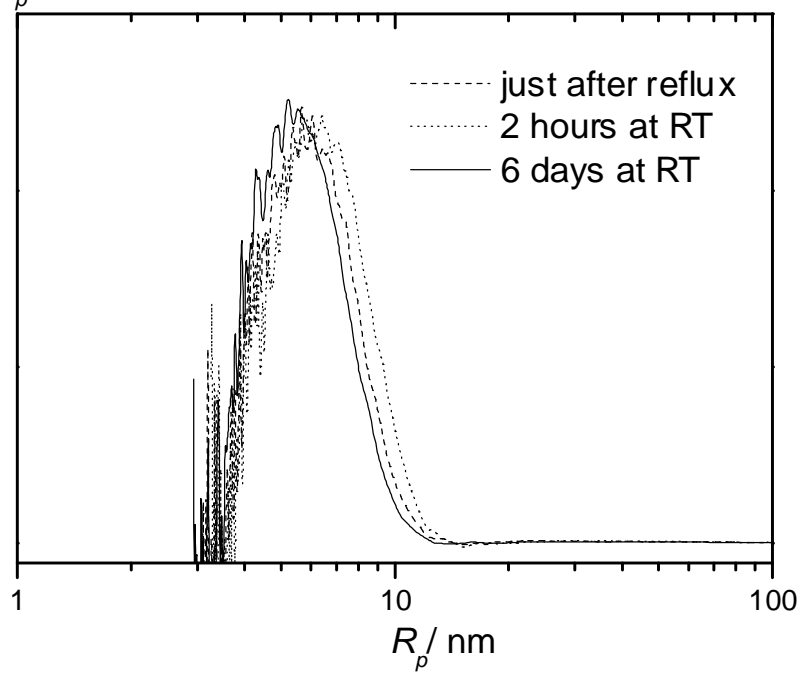

**Figure 6**: Thermograms (**a**) and mesh size distributions (MSD) (**b**) of cross-linked polyethylene sample having received a dose of 1000 kGy and swollen with *o*-xylene. The sample is analyzed, just after the reflux was stopped (dotted line), after two hour resting at room temperature (dashed line) and after six days at room temperature (solid line).

For all the other polymeric samples the same procedure is adopted: two hours reflux followed by 12 hours at room temperature before the thermoporosimetry measurement.

The two chemicals (naphthalene and p-dichlorobenzene), which are solid at room temperature, show an unexpected behavior for polymer samples. While they normally responded in the rigid porous sample cases, no crystallization signal is detected when they are confined in swollen polymeric gels. In both PE and PP/EPR samples, naphthalene and p-dichlorobenzene exhibit melting endothermic peaks corresponding to the confined solvent but no peak of solidification is detected during the cooling stage. A sublimation phenomenon could explain the disappearance of the solvent but this behavior is not observed in the silica or alumina samples.

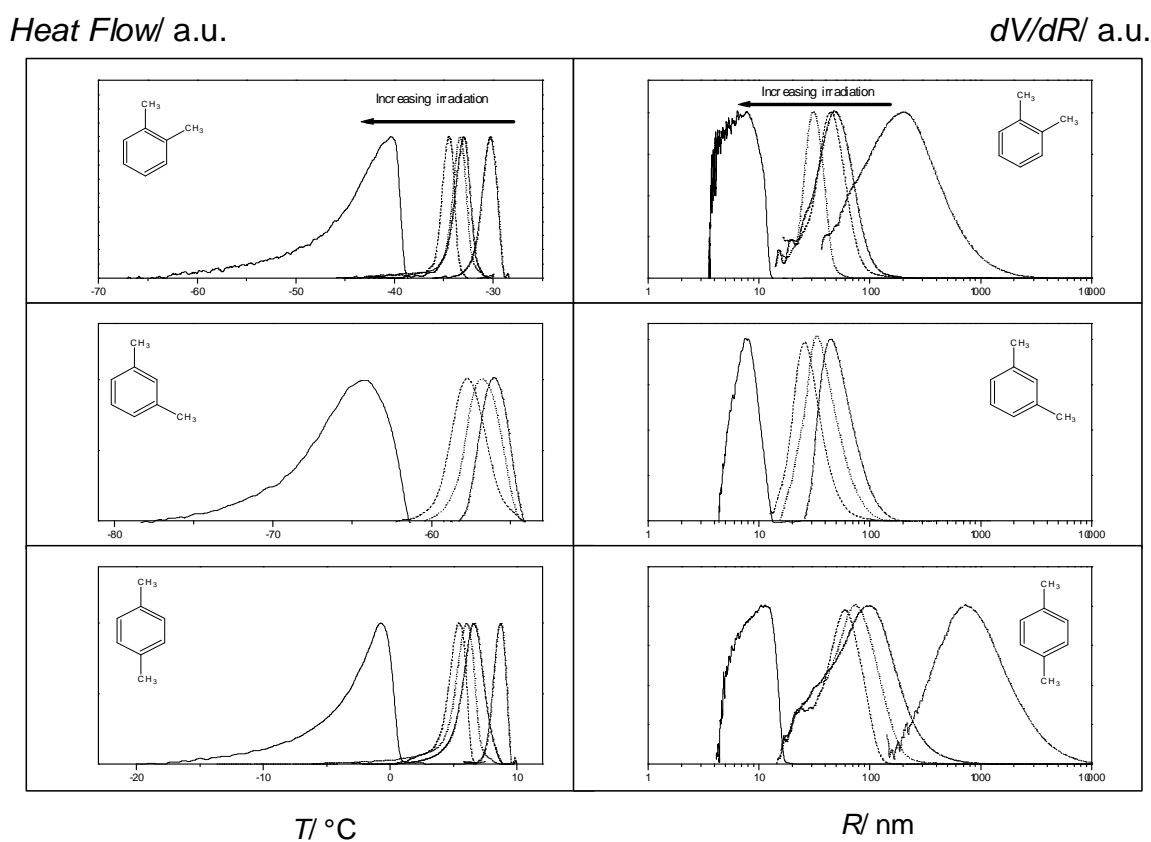

**Figure 7**: Thermograms (on the left) and corresponding mesh size distributions (on the right) of PE1000kGy (solid line), PP940kGy (dotted line), PP830kGy (dashed line), PP620kGy (dash dotted line) and PP405kGy (dash dot dotted line) swollen with *o*-xylene, *m*-xylene and *p*-xylene. The degrees of swelling of these samples in the corresponding solvents are given in the Table 2.

Figure 7 shows the DSC thermal curves and their corresponding transformations in mesh size distributions (MSD) in relation to the five olefins having received different doses of γ-rays. No thermogram was represented for 405kGy/m-xylen sample: the crystallization peak of the confined solvent could not be observed. To make the comparison easier, all the curves are normalized. It is worthy to note that these calculated mesh size distributions are relative to the swollen material and depend on the polymer/solvent pair as well as their mutual interactions namely on their Flory parameter. However, as can be seen in Figure 7, whatever the considered solvent among *o,m* and *p*-xylene, the sequences of the DSC thermal curves (on the left) are the same: the higher the irradiation dose, the more the thermograms are shifted towards lower temperatures. The MSDs (on the right) of these cross linked polyolefins are also in the same order: the more cross-linked the samples, the smaller are the mesh sizes of their swollen networks. A similar agreement is also observed between the swelling degrees (Table 2) of these polymers and their mesh size distributions. This good consistency has to be quantitatively moderated, because most of the MSD needs an extrapolation of the calibration range. However, it remains very indicative for relative analysis. Figure 8 shows the evolution of the maximum of the MSD calculated for the various polymers both as a function of the swelling ratio G (Figure 8 a) and the irradiation dose (Figure 8 b) confirming the previous observations. It is worthy to note that the evolution of $R_{max}$ depends on the nature of the solvent. This could be related to χ the Flory interaction parameter between the polymeric sample and the different solvents. The evolution of G as a function of the irradiation dose is almost linear as shown in Figure 9. From these results it seems that the effect of the solvent which is weak on the value of G is amplified if we look at the size of the meshes ($R_{max}$). This indicates that the MSD is much more sensitive to $\chi$ than G. This technique provides consequently a precious tool to study the thermodynamical quality of a given solvent and could also be used to determine the Flory interaction parameter.

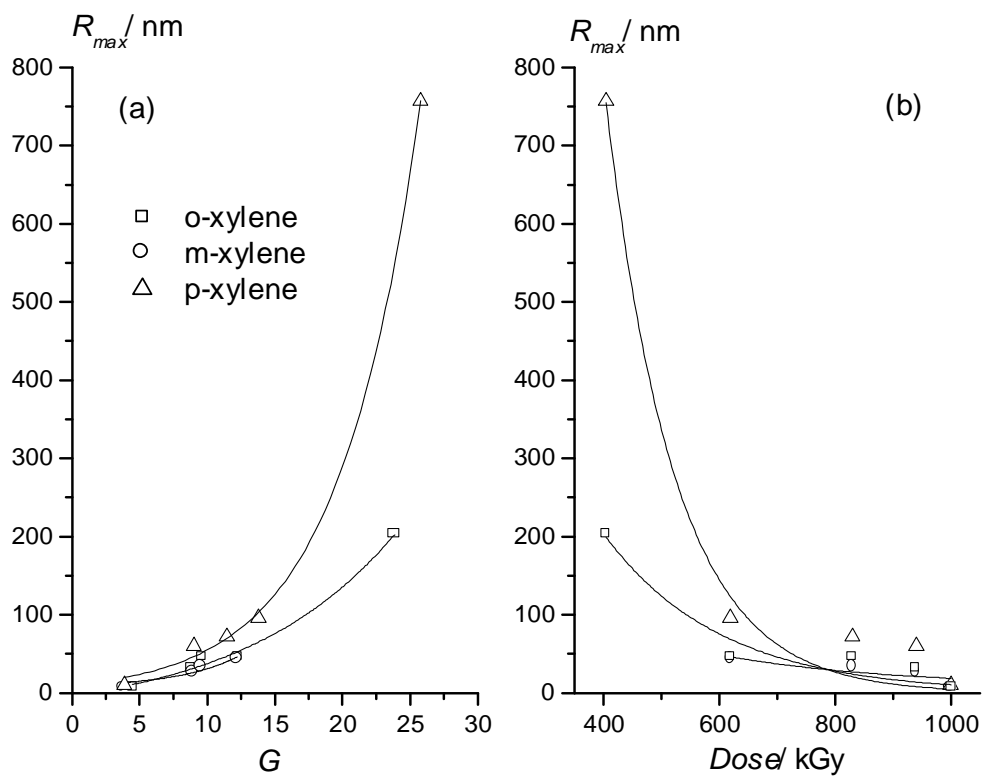

**Figure 8**: Evolution of the maximum of the Mesh Size Distribution ($R_{max}$) for the different polyolefins as a function of the swelling ratio (a) and the irradiation dose (b) for the three xylene isomers. Lines are drawn as a guide for the eye.

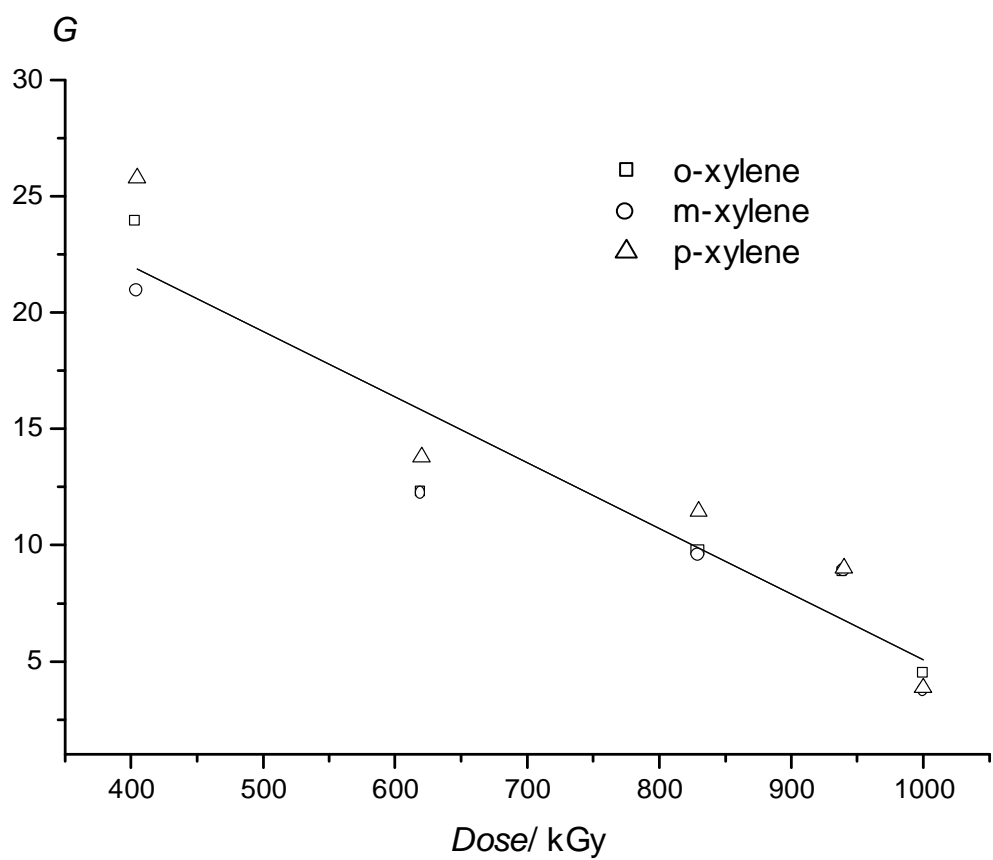

**Figure 9**: Evolution of the degree of swelling G for the three xylene isomers as a function of the irradiation dose received by the polymer. The line corresponding to a linear evolution is drawn as guide for the eye.

# Conclusion

Cross-linked polyethylene and blend of polypropylene/polyethylene were successfully studied by thermoporosimetry analysis. This technique allows the calculation of relative mesh size distribution of swollen polymeric networks giving a direct mean to compare the cross-linking levels of various polyolefins. The mesh size distributions are ordered in the same sequence of that of the degrees of swelling and the γ-radiation doses. Several organic solvents (*o, m, p*-xylene, *p*-dichlorobenzene, *1,2,4*-trichlorobenzene and naphthalene) are calibrated and their numerical equations, needed to conduct all the calculations, are precisely established using porous gel-derived silica samples with well known textural properties.

The main interest of this technique is its suitability even for the heavily dark filled polymers. No special sample preparation is needed provided that the polymer can be swollen by an appropriate solvent. Nevertheless, the mesh size distributions established by this technique are tightly related to the solvent-polymer pair. In addition [15], when the solvent freezes within a polymeric gel, it lead to a stress on the meshes of the network and the topology revealed by the mesh size distribution can differ from the original one. However, the thermoporosimetry remains a helpful tool to compare the cross-linking levels if the same swelling solvent is used.

# Acknowledgements

The authors thank C. Bergeon for kindly providing the PP/EPR samples.

[15] Georges W. Scherer. *Advances in Colloid and Interface Science,* **1998**, *76-77*, 321.

# References

[1] W. Borchard, M. Hostman and W.S. Veeman, *European Polymer Journal,* **2003**, *39*, 1809.

[2] J. Berriot, F. Martin, H. Montes, L. Monnerie and P. Scotta, *Polymer,* **2003**, *44*, 1437.

[3] Roger A. Assink, Kenneth T. Gillen and Briana Sanderson, *Polymer* **2002**, *43*, 1349.

[4] M. Brun, A. Lallemand, J-F. Quinson and C. Eyraud, *Thermochim. Acta* **1977**, *21*, 59.

[5] Y. Saho, G. Hoang, T.W. Zerda, *J. Non-Cryst. Solids,* **1995**, *182*, 309.

[6] K. Ishikiriyama, A. Sakamoto, M. Todoki, T. Tayama, K. Tanaka and T. Kobayashi, *Thermochim. Acta,* **1995,** *267*, 169.

[7] Kazuhiko Ishikiriyama, Minoru Todoki and Kinshi Motomura, *J. Colloid Interf. Sci.,* **1995**, *171*, 92.

[8] ]M.Baba, J.M. Nedelec and J. Lacoste, *J. Phys. Chem. B,* **2003**, *107*, 12884.

[9] M. Iza, S. Woerly, C. Danumah, S. Kaliaguine and M. Bousmina, *Polymer* **2000**, *41*, 5885.

[10] M. Baba, J.M. Nedelec, J. Lacoste, J-L. Gardette and M. Morel, *Polym. Degrad. and Stabil* **2003**, *80*, 305.

[11] M. Baba, J.M. Nedelec, J. Lacoste and J.-L. Gardette, *J. Non-Cryst. Solids,* **2003**, *315*, 228.

[12] M. Brun, J.-F. Quinson, P. Claudy and J.-M Letoffe, *Thermochim. Acta,* **1981**, *44*, 289.

[13] L.L. Hench, in Sol-gel silica: *processing, properties and technology transfer, Noyes Publications, New York,* **1998.**24